\documentstyle[twoside,psfig]{article}
 
\catcode`\@=11
\long\def\@makefntext#1{
\protect\noindent \hbox to 3.2pt {\hskip-.9pt
$^{{\eightrm\@thefnmark}}$\hfil}#1\hfill}               

\def\@makefnmark{\hbox to 0pt{$^{\@thefnmark}$\hss}}    
 
\def\ps@myheadings{\let\@mkboth\@gobbletwo
\def\@oddhead{\hbox{}
\rightmark\hfil\eightrm\thepage}
\def\@oddfoot{}\def\@evenhead{\eightrm\thepage\hfil
\leftmark\hbox{}}\def\@evenfoot{}
\def\sectionmark##1{}\def\subsectionmark##1{}}
 
\oddsidemargin=\evensidemargin
\newcounter{sectionc}\newcounter{subsectionc}\newcounter{subsubsectionc}
\renewcommand{\section}[1] {\vspace{12pt}\addtocounter{sectionc}{1}
\setcounter{subsectionc}{0}\setcounter{subsubsectionc}{0}\noindent
        {\tenbf\thesectionc. #1}\par\vspace{5pt}}
\renewcommand{\subsection}[1] {\vspace{12pt}\addtocounter{subsectionc}{1}
        \setcounter{subsubsectionc}{0}\noindent
        {\bf\thesectionc.\thesubsectionc. {\kern1pt \bfit #1}}\par\vspace{5pt}}
\renewcommand{\subsubsection}[1] {\vspace{12pt}\addtocounter{subsubsectionc}{1}
        \noindent{\tenrm\thesectionc.\thesubsectionc.\thesubsubsectionc.
        {\kern1pt \tenit #1}}\par\vspace{5pt}}
\newcommand{\nonumsection}[1] {\vspace{12pt}\noindent{\tenbf #1}
        \par\vspace{5pt}}
 
\newcounter{appendixc}
\newcounter{subappendixc}[appendixc]
\newcounter{subsubappendixc}[subappendixc]
\renewcommand{\thesubappendixc}{\Alph{appendixc}.\arabic{subappendixc}}
\renewcommand{\thesubsubappendixc}
        {\Alph{appendixc}.\arabic{subappendixc}.\arabic{subsubappendixc}}

\renewcommand{\appendix}[1] {\vspace{12pt}
        \refstepcounter{appendixc}
        \setcounter{figure}{0}
        \setcounter{table}{0}
        \setcounter{lemma}{0}
        \setcounter{theorem}{0}
        \setcounter{corollary}{0}
        \setcounter{definition}{0}
        \setcounter{equation}{0}
        \renewcommand{\thefigure}{\Alph{appendixc}.\arabic{figure}}
        \renewcommand{\thetable}{\Alph{appendixc}.\arabic{table}}
        \renewcommand{\theappendixc}{\Alph{appendixc}}
        \renewcommand{\thelemma}{\Alph{appendixc}.\arabic{lemma}}
        \renewcommand{\thetheorem}{\Alph{appendixc}.\arabic{theorem}}
        \renewcommand{\thedefinition}{\Alph{appendixc}.\arabic{definition}}
        \renewcommand{\thecorollary}{\Alph{appendixc}.\arabic{corollary}}
        \renewcommand{\theequation}{\Alph{appendixc}.\arabic{equation}}
        \noindent{\tenbf Appendix \theappendixc #1}\par\vspace{5pt}}
\newcommand{\subappendix}[1] {\vspace{12pt}
        \refstepcounter{subappendixc}
        \noindent{\bf Appendix \thesubappendixc. {\kern1pt \bfit #1}}
        \par\vspace{5pt}}
\newcommand{\subsubappendix}[1] {\vspace{12pt}
        \refstepcounter{subsubappendixc}
        \noindent{\rm Appendix \thesubsubappendixc. {\kern1pt \tenit #1}}
        \par\vspace{5pt}}
 
\newcommand{\textlineskip}{\baselineskip=13pt}
\newcommand{\smalllineskip}{\baselineskip=10pt}
 
\def\eightcirc{
\begin{picture}(0,0)
\put(4.4,1.8){\circle{6.5}}
\end{picture}}
\def\eightcopyright{\eightcirc\kern2.7pt\hbox{\eightrm c}}
\newcommand{\copyrightheading}[1]
        {\vspace*{-2.5cm}\smalllineskip{\flushleft
        {\footnotesize Modern Physics Letters A #1}\\
        {\footnotesize $\eightcopyright$\, World Scientific Publishing
         Company}\\
         }}
 
\newcommand{\publisher}[2]{{\begin{center}\footnotesize\smalllineskip
        Received #1\\
        Revised #2
        \end{center}
        }}
 
\def\abstracts#1#2#3{{
        \centering{\begin{minipage}{4.5in}\footnotesize\baselineskip=10pt
        \parindent=0pt #1\par
        \parindent=15pt #2\par
        \parindent=15pt #3
        \end{minipage}}\par}}
 
\newcommand{\bibit}{\nineit}
\newcommand{\bibbf}{\ninebf}
\renewenvironment{thebibliography}[1]
        {\frenchspacing
         \ninerm\baselineskip=11pt
         \begin{list}{\arabic{enumi}.}
        {\usecounter{enumi}\setlength{\parsep}{0pt}
         \setlength{\leftmargin 12.7pt}{\rightmargin 0pt} 
         \setlength{\itemsep}{0pt} \settowidth
        {\labelwidth}{#1.}\sloppy}}{\end{list}}
 
\newcounter{itemlistc}
\newcounter{romanlistc}
\newcounter{alphlistc}
\newcounter{arabiclistc}

\newcommand{\fcaption}[1]{
        \refstepcounter{figure}
        \setbox\@tempboxa = \hbox{\footnotesize Fig.~\thefigure. #1}
        \ifdim \wd\@tempboxa > 5in
           {\begin{center}
        \parbox{5in}{\footnotesize\smalllineskip Fig.~\thefigure. #1}
            \end{center}}
        \else
             {\begin{center}
             {\footnotesize Fig.~\thefigure. #1}
              \end{center}}
        \fi}
 
\newcommand{\tcaption}[1]{
        \refstepcounter{table}
        \setbox\@tempboxa = \hbox{\footnotesize Table~\thetable. #1}
        \ifdim \wd\@tempboxa > 5in
           {\begin{center}
        \parbox{5in}{\footnotesize\smalllineskip Table~\thetable. #1}
            \end{center}}
        \else
             {\begin{center}
             {\footnotesize Table~\thetable. #1}
              \end{center}}
        \fi}
 
\def\@citex[#1]#2{\if@filesw\immediate\write\@auxout
        {\string\citation{#2}}\fi
\def\@citea{}\@cite{\@for\@citeb:=#2\do
        {\@citea\def\@citea{,}\@ifundefined
        {b@\@citeb}{{\bf ?}\@warning
        {Citation `\@citeb' on page \thepage \space undefined}}
        {\csname b@\@citeb\endcsname}}}{#1}}
 
\newif\if@cghi
\def\cite{\@cghitrue\@ifnextchar [{\@tempswatrue
        \@citex}{\@tempswafalse\@citex[]}}
\def\citelow{\@cghifalse\@ifnextchar [{\@tempswatrue
        \@citex}{\@tempswafalse\@citex[]}}
\def\@cite#1#2{{$\null^{#1}$\if@tempswa\typeout
        {IJCGA warning: optional citation argument
        ignored: `#2'} \fi}}

                                                                                \def\pmb#1{\setbox0=\hbox{#1}
        \kern-.025em\copy0\kern-\wd0
        \kern.05em\copy0\kern-\wd0
        \kern-.025em\raise.0433em\box0}

\def\fnt#1#2{\footnotetext{\kern-.3em
        {$^{\mbox{\scriptsize #1}}$}{#2}}}
 
\def\ps@myheadings{%
    \let\@oddfoot\@empty\let\@evenfoot\@empty
    \def\@evenhead{\slshape\leftmark\hfil}
    \def\@oddhead{\hfil{\slshape\rightmark}}
    \let\@mkboth\@gobbletwo
    \let\sectionmark\@gobble
    \let\subsectionmark\@gobble
    }
\font\tenrm=cmr10
\font\tenit=cmti10
\font\tenbf=cmbx10
\font\bfit=cmbxti10 at 10pt
\font\ninerm=cmr9
\font\nineit=cmti9
\font\ninebf=cmbx9
\font\eightrm=cmr8

\def\qed{\hbox{${\vcenter{\vbox{                        
   \hrule height 0.4pt\hbox{\vrule width 0.4pt height 6pt
   \kern5pt\vrule width 0.4pt}\hrule height 0.4pt}}}$}}
 
\begin{document}
\setlength{\textheight}{7.7truein}  
 
\thispagestyle{empty}
 
\markboth{\protect{\footnotesize\it Instructions for Typesetting
Manuscripts}}{\protect{\footnotesize\it Instructions for
Typesetting Manuscripts}}
 
\normalsize\textlineskip
 
\setcounter{page}{1}
 
\copyrightheading{}     
 
\vspace*{0.88truein}

 \centerline{\bf LANDAU--ZENER EFFECT }
\baselineskip=13pt
\centerline{\bf IN SUPERFLUID NUCLEAR SYSTEMS}
\vspace*{0.4truein}
\centerline{\footnotesize M. MIREA\footnote{
email: mirea@ifin.nipne.ro
}}
\baselineskip=12pt
\centerline{\footnotesize\it Nuclear Physics Department, Institute of
Physics and Nuclear Engineering, P.O. Box MG--6
}
\baselineskip=10pt
\centerline{\footnotesize\it Bucharest, Romania}
\vspace*{12pt}

\publisher{(received date)}{(revised date)}
 
\vspace*{0.23truein}
\abstracts{The Landau--Zener effect is generalized for many-body systems
with pairing residual interactions.  The microscopic equations of
motion are obtained and the $^{14}$C decay of $^{223}$Ra  spectroscopic
factors are deduced.
An asymmetric nuclear shape parametrization given by two intersected
spheres is used. The single particle level scheme is determined in the
frame of the superasymmetric two-center shell. The deformation energy
is computed in the microscopic-macroscopic approximation. The
penetrabilities are obtained within the WKB approximation.
The fine structure of the cluster decay
analyzed in the frame of this formalism
gives a very good agreement with the experimental ratio of partial half-lives
obtained in special conditions.
}{}{}
 

\vspace*{2pt}
 
 
\baselineskip=13pt              
\normalsize                     
\section{Introduction}          
\vspace*{-0.5pt}
\noindent
Recently, a new method was proposed to approach the fine structure
phenomenon of cluster- and alpha-decays in odd-nuclei.
These decays modes are considered as 
superasymmetric fission
processes. Assuming the existence of few collective variables, associated
to some generalized coordinates describing the nuclear shape
parametrization, this approach allows to
handle approximately the behavior of many other intrinsic
variables. 
At any values of the shape generalized coordinates, 
the single particle level scheme
and the potential barrier are determined. During the whole process and
after the disintegration, the  single particle occupation
probabilities of the orbitals of interest 
are computed using the Landau-Zener promotion mechanism alone. This
method was successfully used to describe the $^{14}$C- and alpha-decays 
\cite{mm1,mm2,mm3}.
Despite the overlook of residual interactions, the agreement 
obtained with data gives a strong experimental
support to the formalism. 
Unfortunately, the approach based on the Landau-Zener effect alone
is not able to describe the fine structure in the case of the
disintegration of even-even nuclei.
That motivated us to develop the model in order
to actually obtain the final occupation probabilities 
of the nucleons in different orbitals using the equation of motion derived for
pairing residual interactions including the Landau-Zener effect.

The decaying system provides a time dependent single-particle 
potential in which the nucleons move independently. This description
is consistent within the essence of the Hartree-Fock approximation
where the many body wave function is constrained to be Slater
determinants and with the Hartree-Fock-Bogoliubov (HFB) problem
where the function is constrained to be of BCS form all time.
In the HFB model, the level slippage is treated automatically
due to the simultaneous presence of many Slater determinants in
the wave function. Such calculations are very cumbersome. A way to
bypass this problem is to use the superfluid model constructed over
a single particle potential and introducing the Landau-Zener effect
to replace the effect given by a part of the residual interactions.

In the next section, the equations describing the microscopic
dynamics of a many nucleon system with pairing residual
interaction, including the Landau-Zener effect applied to the
unpaired nucleon, are deduced.
This effect allows the transition of an unpaired nucleon between
levels with the same quantum numbers associated to some symmetries
of the system. Solving the equations, the response of the system
is estimated when the nuclear shape parametrization is changed.
In the following, a nuclear shape parametrization
given only by two intersected spheres, the radius of the sphere
associated to the light fragment kept as constant, will be used.
The single degree of freedom remains the elongation characterized
by $R$, the distance between the centers of the nascent fragments.
Sometimes a normalized elongation 
$(R-R_{i})/(R_{f}-R_{i})$ will be used for convenience.
The parameter $R_{1}$ being the radius of the daughter 
determined from the condition of volume conservation and 
$R_{2}=41A_{2}^{1/3}$ being the radius of the light fragment kept as constant, 
the next values are determined
$R_{i}=R_{1}-R_{2}$ and $R_{f}=R_{1}+R_{2}$ for the configuration of
the initial nucleus considered as a sphere and for that
of two tangent spheres, respectively.

\section{Formalism}          
\noindent
In order to obtain the equations of motion, we shall 
start from the variational principle taking the 
Lagrangian as
\begin{equation}
{\cal L}=<\phi \mid H-i\hbar{\partial 
\over \partial t}+H'-\lambda N\mid \phi>
\end{equation}
and assuming the many-body state formally expanded as a superposition of 
time dependent BCS seniority one diabatic wave functions
\begin{equation}
\mid \phi>=\sum_{i} c_{i}(t)a_{i}^{+}\prod_{j\ne i}
(u_{j}(t)+v_{j}(t)a_{j}^{+}a_{\bar{j}}^{+})\mid 0>
\end{equation}
Each of these seniority one diabatic wave functions being associated to
a state with an unpaired nucleon in the orbital $i$, the orbitals $j\ne i$
being pairwise filled. 
Furthermore
\begin{equation}
H(t)=\sum_{k>0}\epsilon_{k}(t)(a_{k}^{+}a_{k}+a_{\bar {k}}^{+}a_{\bar {k}})
-G\sum_{k,l>0}a_{k}^{+}a_{\bar {k}}^{+}a_{\bar{l}}a_{l}
\end{equation}
is the time dependent many-body Hamiltonian with pairing residual
interactions,
\begin{equation}
\begin{array}{c}
H'={1\over 2}\sum_{i,j}h_{ij}(t)[(u_{i}a_{i}-v_{i}a_{i}^{+})
(u_{j}^{*}a_{j}^{+}+v_{j}^{*}a_{\bar{j}})+(u_{j}a_{j}-v_{j}a_{j}^{+})
(u_{i}^{*}a_{i}^{+}+v_{i}^{*}a_{\bar{i}})]
\end{array}
\end{equation}
is the residual interaction between diabatic levels 
characterized by the same quantum
numbers associated to some symmetries of the system.
This interaction is responsible for the existence of avoiding
level crossing levels in the case of adiabatic states. 
The presence of this interaction  term $H'$
allows the level slippage of an unpaired nucleon
and allows the possibility that 
a unpaired nucleon (or hole) located on the diabatic level $i$ jumps on the
level $j$ due to the interaction $h_{ij}(t)$, in the same manner as in
the classical two level Landau--Zener effect \cite{sch1,sch2}. The element
$h_{ij}(t)$ is different from zero only in the regions of the
avoided crossings. 
The term $-\lambda N$ represents 
the well known constraint on the total
number of particles. 
In this picture, $a_{i}^{+}$ and $a_{i}$ denote
operators for creating and destroying, respectively, a particle in
the diabatic state $i$ (not, as usual, in the adiabatic state $i$) and
the label $\bar{i}$ denotes the time-reversed orbital conjugate to,
and degenerate with, $i$. Furthermore, $\epsilon_{k}$ are the
single particle energies obtained within the shell model.
The blocking correlations (the variation in $\Delta$ and in the
$u_{j}$'s and $v_{j}$'s due to the change of the blocking level\cite{ring})
are neglected.

By performing the variation of the Lagrangian in a way similar as
in Ref \cite{blo}, the next system of 
coupled differential equations are obtained.
\begin{equation}
i\hbar\dot{\rho}_{l}=\sum_{m}p_{m}
\{ \kappa_{l}\Delta_{m}^{*}-\kappa_{l}^{*}\Delta_{m}\}
\label{e5}
\end{equation}
\begin{equation}
i\hbar\dot{\kappa}_{l}=\sum_{m}p_{m}
\{(2\rho_{l}-1)\Delta_{m}-2\kappa_{l}(\epsilon_{l}(t)-\lambda(t))\}
\label{e6}
\end{equation}
\begin{equation}
i\hbar\dot{p}_{m}={1\over 2}\sum_{j}h_{mj}(t)(S_{mj}-S_{jm})
\label{e7}
\end{equation}
\begin{equation}
\begin{array}{c}
i\hbar\dot{S}_{jm}=S_{jm}\{-{1\over G}(\mid \Delta_{m}\mid^{2}-
\mid \Delta_{j}\mid^{2})+(\epsilon_{m}(t)-\epsilon_{j}(t))-\\ 
{1\over 2}[-{\rho_{m}\over \kappa_{m}}
+2\kappa_{m}^{*}+{\rho_{j}\over \kappa_{j}}-2\kappa_{j}^{*}]
\sum_{l}p_{l}\Delta_{l}-\\
{1\over 2}[-{\rho_{m}\over \kappa_{m}^{*}}
+2\kappa_{m}+{\rho_{j}\over \kappa_{j}^{*}}-2\kappa_{j}]
\sum_{l}p_{l}\Delta_{l}^{*}\}+
{1\over 2}\sum_{k\ne m,j}(h_{mk}(t)S_{jk}-
h_{jk}(t)S_{km})+
{1\over 2}h_{jm}(t)(p_{j}-p_{m})
\end{array}
\label{e8}
\end{equation}
where
\begin{equation}
\begin{array}{c}
p_{m}=c_{m}^{*}c_{m}\\
S_{jm}=c_{j}^{*}c_{m}\\
\Delta_{m}=G\sum_{k\ne m}\kappa_{k}\\
\Delta_{m}^{*}=G\sum_{k\ne m}\kappa_{k}^{*}\\
\kappa_{k}=u_{k}^{*}v_{k}\\
\rho_{k}=\mid v_{k}\mid^{2}
\end{array}
\end{equation}
In this description, $p_{m}$ represents the single particle
occupation probability of the orbital $m$, and $\rho_{k}$ is
the pairwise occupation probability of the orbital $k$.
The previous system is a generalization of the equations of
motions obtained in Refs. \cite{blo,koo,neg}  
and are
applied in the case of fine structure of the $^{14}$C emission from $^{223}$Ra.
This is a rare phenomenon where the theory \cite{mg,rus} has superseded 
the experiments \cite{ho1,ho2}. However, the predictions were not able to assess
that transitions to the first excited state are favored.
The system 
is solved by taking as initial values those corresponding
to the ground state $1i_{11/2}$ of the parent $^{223}$Ra. 
The starting elongation is that which
gives the minimal value of the microscopic-macroscopic deformation energy.

\section{Results Concerning the $^{14}$C Decay of $^{223}$Ra}
\noindent

\begin{figure}[htbp] 
\vspace*{13pt}
 \centerline{\psfig{file=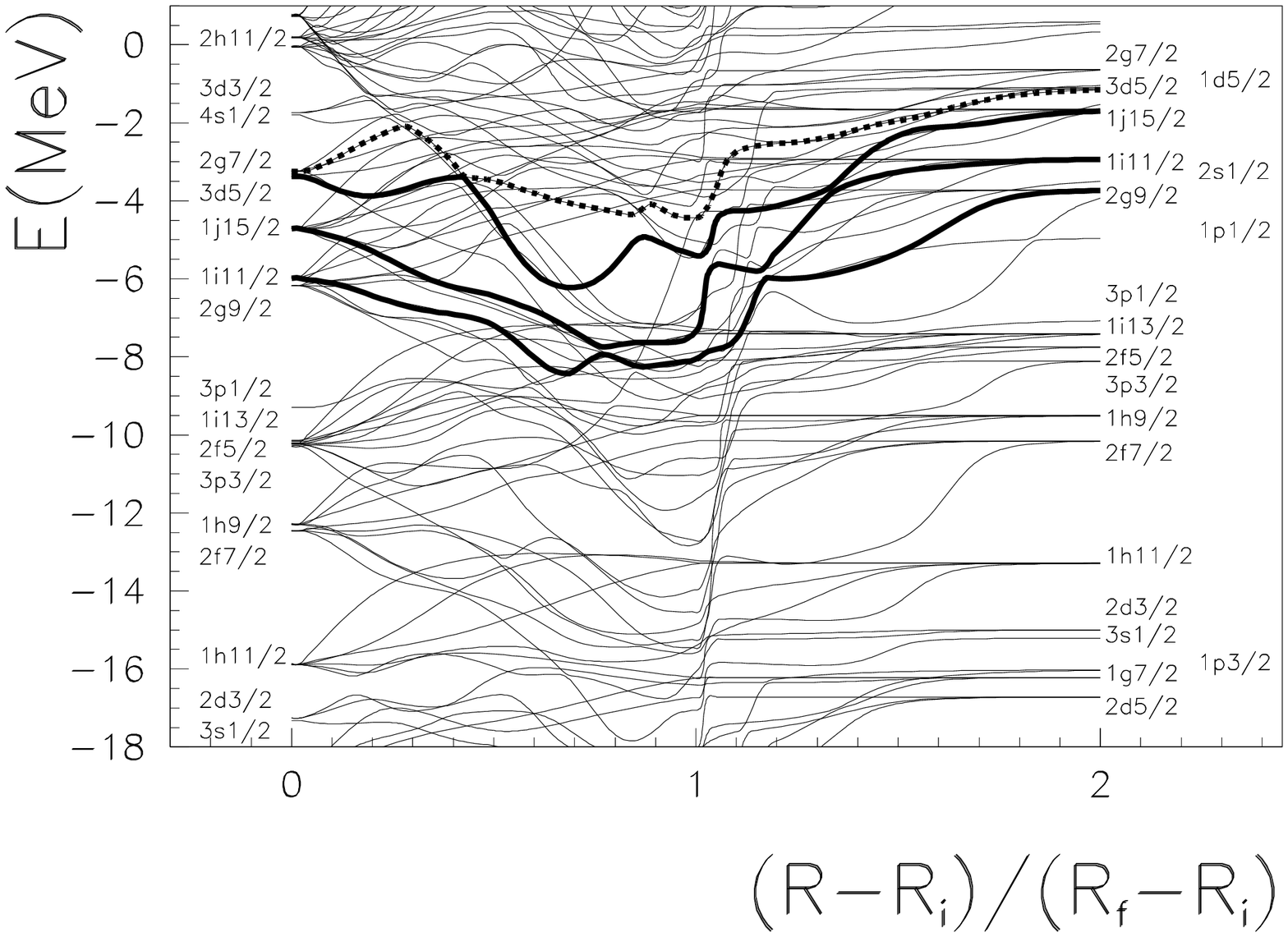,width=12cm}} 
\fcaption{Neutron level scheme for $^{14}$C spontaneous emission from
$^{223}$Ra with respect the normalized elongation. The levels 
(with $\Omega$=3/2) emerging from
1$i_{11/2}$, 1$j_{15/2}$ and 3$d_{5/2}$ are represented with thick
lines. }
\vspace*{13pt}
\label{fig1}
\end{figure}

The neutron and proton level schemes are obtained with the 
superasymmetric two-center shell model \cite{mm5} improved in
Ref. \cite{mm1}. The single particle neutron levels are
displayed in Fig. \ref{fig1}. The $^{223}$Ra has the spin ${3\over 2}$
emerging from 1$i_{11/2}$. Adiabatically, this unpaired neutron
reaches the 2$g_{9/2}$ level of the daughter $^{209}$Pb. In the
frame of this formalism, the fine structure of the $^{14}$C radioactivity
can be understood by an enhanced transition probability of the
unpaired neutron from the adiabatic level with $\Omega=3/2$
emerging from 1$i_{11/2}$, to the adiabatic level with the same
spin projection emerging from $1j_{15/2}$. The level scheme
presented in Fig. \ref{fig1} shows that the 1$i_{11/2}$ level
reaches adiabatically the 2$g_{9/2}$ state, the $1j_{15/2}$ reaches the
1$i_{11/2}$ state, and the 3$d_{5/2}$ level reaches 1$j_{15/2}$ state of the
daughter $^{209}$Pb. In system with cylindrical symmetries, Landau-Zener
transitions
can be realized only between levels with the same spin projection $\Omega$. 
Four avoided level crossings can be observed in Fig. \ref{fig1} produced
at $(R-R_{i})/(R_{f}-R_{i})\approx$ 0.75, 1, 1.1 and 1.2. Without these
interactions, diabatically, the levels $1i_{11/2}$, $1j_{15/2}$ and $3d_{5/2}$
attain \cite{mm1} the 1$i_{11/2}$, $1j_{15/2}$ and $2g_{9/2}$ 
daughter orbitals, respectively.

In Ref. \cite{koo}, a particular form of the equations (\ref{e5}-\ref{e8}) 
(without Landau-Zener included) were used to study the 
dumping (or the dissipation) process. The dissipation was
defined as the flow of energy between collective and intrinsic
modes. This energy flow must be irreversible. Unfortunately,
accordingly with the discussion of Ref. \cite{koo},
neglecting the Landau-Zener effect, the calculation is microscopically
reversible in the sense that if all the coordinates are time-reversed
the system retraces its path. An unknown fraction of the energy that
is identified as dissipated energy is in fact collective kinetic energy.
Through the presence of the Landau-Zener term
in the equations of motion the requirement of
irreversibility is satisfied. Therefore, The Landau-Zener probabilities 
$p_{m}$ describe a true dissipation process, while the variations
of $\rho_{k}$ and $\kappa_{k}$ values take partially into account
the collective kinetic energy. In these circumstances, it can be 
assumed that, after the disintegration, the spectroscopic factor
can be approximated with only the leading term:
\begin{equation}
S_{m}=\mid <\Phi\mid\Phi_{0m}>\mid^{2}\propto p_{m}
\end{equation}
where $\Phi$ is the BCS wave function given by the equations of motion
after the scission, $\Phi_{0m}=a_{m}^{+}\prod_{j\ne m}(u_{0j}+v_{0j}
a_{j}^{+}a_{\bar{j}}^{+})\mid 0>$, with 
$m$=$2g_{9/2}$, $1i_{11/2}$, $1j_{15/2}$,
 are the orthogonal set of seniority one wave functions constructed
on the asymptotic overlapped energy levels of the daughter and the
emitted nucleus with the ground state $u_{0j}$ and $v_{0j}$ values.

\begin{figure}[htbp] 
\vspace*{13pt}
\centerline{\psfig{file=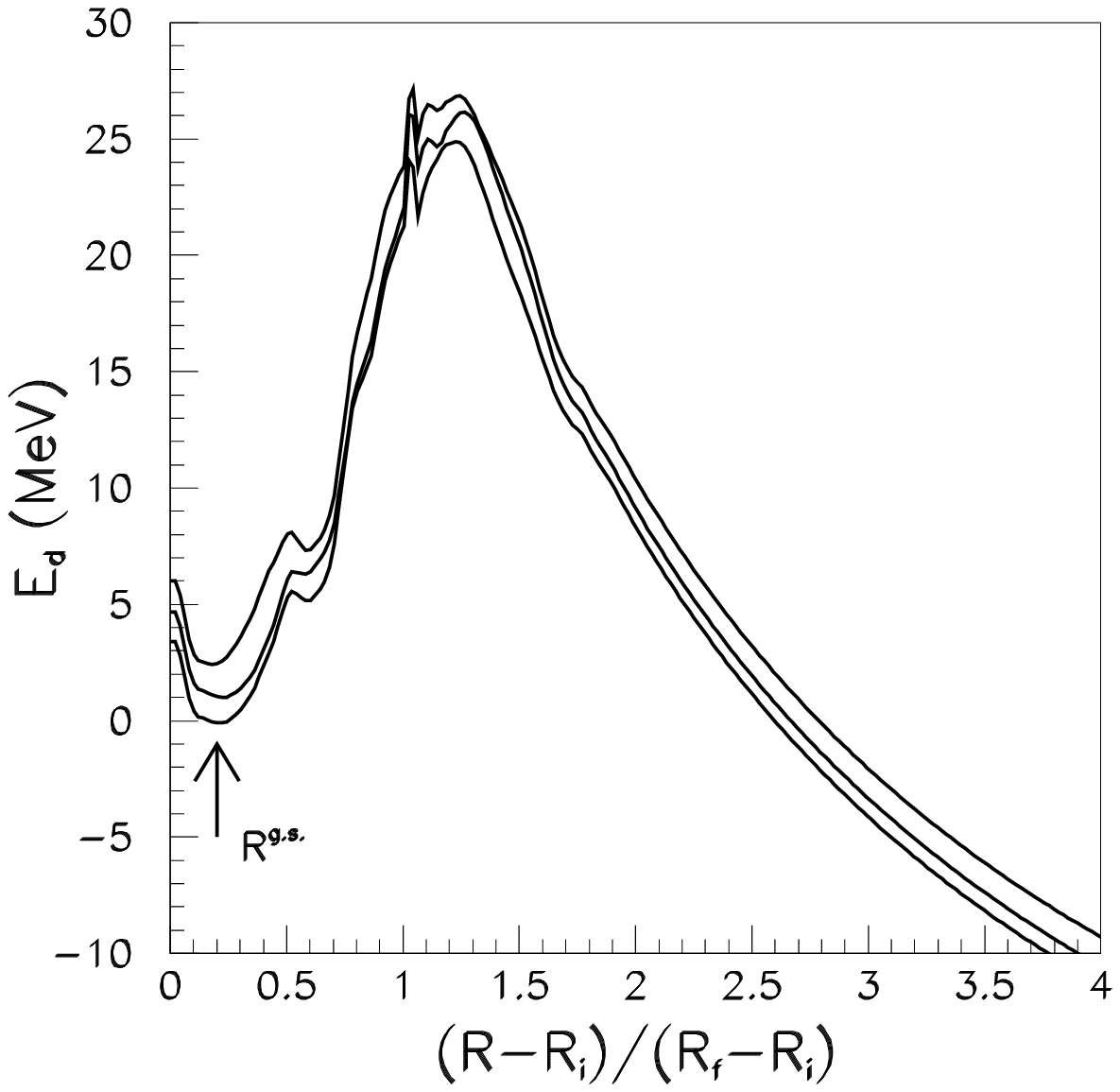,width=12cm}} 
\vspace*{13pt}
\fcaption{The potential barrier $E_{d}$ measured from the ground state of the
parent as function of the normalized elongation coordinate
$(R-R_{i})/(R_{f}-R{i})$. 
The barriers of the first and second excited states obtained by adding
the single particle excitation energy are also plotted. }
\label{fig2}
\end{figure}

\begin{figure}[htbp] 
\vspace*{13pt}
\centerline{\psfig{file=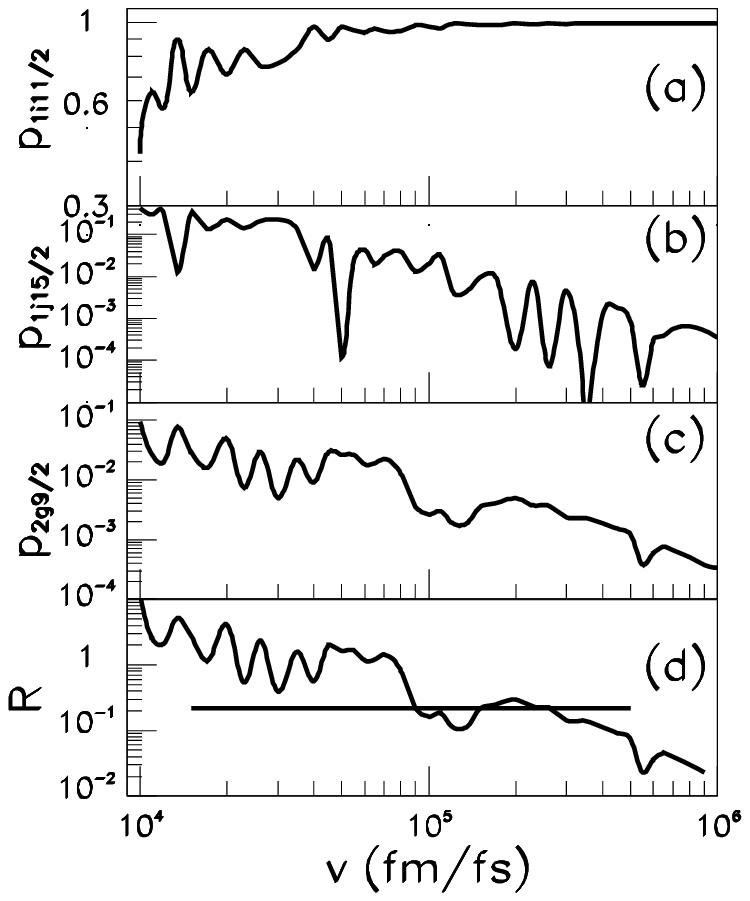,width=12cm}} 
\fcaption{ The single particle occupation probabilities $p$ of the
daughter $^{209}$Pb. Bottom, the ratio $R$ between the partial half-lives for 
transitions to the ground state and first excited state. A full line indicate
the experimental $R$-value. 
 }
\vspace*{13pt}
\label{fig3}
\end{figure}

The partial half-life for the state $i$ is obtained with
the equation in the WKB approximation:
\begin{equation}
T_{1/2}^{m}={h \ln 2\over 2E_{v}p_{m}}\exp\left({2\over
\hbar}\int_{R^{\rm g.s.}}^{R^{m}}
\sqrt{2{A_{1}A_{2}\over A_{0}}E_{d}^{m}}{\rm d}R\right)=
{h \ln 2\over 2E_{v}p_{m}}\exp(K_{m})
\end{equation}
where $p_{m}$ is the final occupation probability and has
the meaning of a spectroscopic factor. $E_{d}^{m}$ is the deformation
energy computed in the framework of the Yukawa-plus-exponential model
\cite{davies} extended for binary systems with different charge densities
\cite{poe}. This deformation energy is corrected within
shell effect computed within Strutinsky's method as in Ref. \cite{mm4}.
The deformation energies for the excited states are obtained
by adding the difference between the single particle energy of the
excited state and the single particle energy of the nucleon in the
ground state. The values of $K_{i}$ are 2.10$\times 10^{27}$,
1.27$\times 10^{29}$ and 1.87$\times 10^{31}$ for transitions to the
daughter ground state, first excited state and 
second excited state, respectively.
These deformation energies are  plotted with thick line
in Fig. \ref{fig2}. $R^{\rm g.s.}$ is the ground state elongation
used as starting point of the decay,
$R^{i}$ is the channel exit point from the barrier and $E_{v}$ is the
zero point vibration energy. 
The effort is mainly focused to reproduce the experimental ratio $R$ of the partial 
half-lives for transitions to the first excited state and to ground state,
\begin{equation}
R={T_{1/2}^{i_{11/2}}\over T_{1/2}^{ g_{9/2}}}=
{p_{g_{ 9/2}}\over p_{i_{11/2}}}{\exp(K_{i_{11/2}})\over \exp(K_{g_{ 9/2}})}
\end{equation}
which has approximately the value 0.218. A critical parameter in the model
is the quantity which characterizes the variations in time of the generalized coordinates.
In the present work, it is matter of the velocity $v$  of the inter-nuclear distance
(or the elongation). Some models are able to determine this quantity \cite{ca}, 
but in the present work the velocity of the elongation is considered as a fitting
parameter and its optimal value will be briefly discussed in comparison with 
previous results. In Fig. \ref{fig3}, the final single-particle occupation 
probabilities as function of the internuclear distance velocity are plotted.
As expected, for large velocities the system becomes 
mainly diabatic and the occupation
probabilities of the first daughter excited state tends to 1 while that of the
ground state reaches 0. For lower velocities, the g.s. transitions are favored. In the
bottom of Fig. \ref{fig3}, the ratio $R$ is plotted. When 
$v\approx 9\times 10^{4}-3\times 10^{5} $, $R$ reaches the vicinity of
the experimental value. The value of $v$ determined in this way is 
approximately one order of magnitude
 lower than that 
determined from the tunneling times of Ref. \cite{ca} using macroscopic models, that means,
approximating the effective mass with a value close to the reduced mass.
However, in the region of avoided crossing levels,  the more realistic 
cranking approximation predicts that the inertia increase with orders of magnitude.
If the effective mass increases, the deformation velocity must decrease in order
to conserve the available energy. Therefore, the lower values of $v$ are
qualitatively consistent with the use of the cranking model for the inertia.

The previous treatment provides a different way to attack the fine structure
of superasymmetric fission for odd and even nuclei. To our knowledge, it
is the unique model which can explain the favored transitions to the
excited state in the case of cluster-decay. Also, the inclusion
of the Landau-Zener effect in the equations of motion offers a more complete
description of dumping phenomena during the disintegration.

\nonumsection{Acknowledgments}
\noindent
This work was partially sponsored by the European Excellence Center IDRANAP.

\nonumsection{References}


\noindent


\begin{thebibliography}{000}

\bibitem{mm1}
M. Mirea, {\bibit Phys. Rev.} {\bibbf C57}, 2484 (1998).

\bibitem{mm2}
M. Mirea, {\bibit Europ. Phys. J.} {\bibbf A4}, 335 (1999).

\bibitem{mm3}
M. Mirea, {\bibit Phys. Rev.} {\bibbf C63}, 034603 (2001).

\bibitem{sch1}
J.Y. Park, W. Greiner and W. Scheid, {\bibit Phys. Rev.} 
{\bibbf C21}, 958 (1980).

\bibitem{sch2} Moon Hoe Cha, J.Y. Park and W. Scheid
{\bibit Phys. Rev.} {\bibbf C36}, 2341 (1987).

\bibitem{ring} P. Ring and P. Schuck, {\bibit The Nuclear
Many--Body Problem} (Springer Verlag, 1980), chap. 6, pp. 238.

\bibitem{blo}
J. Blocki and H. Flocard, {\bibit Nucl. Phys.} {\bibbf A273}, 45 (1976).

\bibitem{koo}
S.E. Koonin and J.R. Nix, {\bibit Phys. Rev.} {\bibbf C13}, 209 (1976).

\bibitem{neg}
J.W. Negele, S.E. Koonin, P. M\"{o}ller, J.R. Nix and A.J. Sierk,
{\bibit Phys. Rev.} {\bibbf C17}, 1098 (1978). 

\bibitem{mg} M. Greiner and W. Scheid, {\bibit J. Phys. G} {\bibbf 12}, L229 (1986).

\bibitem{rus} V.P. Bugrov, S.G. Kadmenskii, V.I. Furman and Yu.M. Chuvilskii,
{\bibit Nuclear Spectroscopy and Nuclear Structure} (in Russian, Nauka, Leningrad, 1987),
pp. 439.

\bibitem{ho1} L. Brillard, A.G. El Ayi, E. Hourani, M. Hussonnois, J.F. Le Du,
L.H. Rosier and L Stab, {\bibit  C.R. Acad. Sci. Paris} {\bibbf 39}, ser. 2, 1105
(1989).

\bibitem{ho2} E. Hourany, G. Berrer-Ronsin, A. Elayi, P. Hoffmann-Rothe, A.C. Mueller,
L. Rosier, G. Rotbard, G. Renou, A. Liebe, D.N. Poenaru and H.L. Ravn,
{\bibit Phys. Rev.} {\bibbf C52}, 267 (1995).


\bibitem{davies}
K.T.R. Davies and J.R. Nix, {\bibit Phys. Rev.} {\bibbf C14},
1977 (1976).

\bibitem{poe}
D.N. Poenaru, M. Ivascu and D. Mazilu, {\bibit Comput. Phys. Comm}
 {\bibbf 19}, 205 (1980).

\bibitem{mm4}
M. Mirea, O. Bajeat, F. Clapier, F. Ibrahim, A.C. Mueller, N. Pauwels
and J. Proust,
{\bibit Europ. Phys. J.} {\bibbf A11}, 59 (2001).

\bibitem{mm5} 
M. Mirea, {\bibit Phys. Rev.} {\bibbf C54}, 302 (1996). 

\bibitem{ca}
N. Carjan, O. Serot and D. Strottman, {\bibit Z. Phys. A} {\bibbf 349}, 353 (1994).
\end{thebibliography}
\end{document}